# Status of the CTA medium size telescope prototype

J.Baehr[a] for the CTA consortium

[a]DESY, Platanenallee 6, D-15745 Zeuthen, Ger many

**Abstract.** We present here the status of the medium size prototype for the Cherenkov Telescope Array. The main reasons to build the prototype are the test of the steel structure, the training of various mounting operations, the test of the drive system and the test of the safety system. The essential difference between the medium size telescope prototype and a fully instrumented are that the camera is not instrumented and only a part of the mounted mirrors are optical mirrors. Insofar no high energy gamma rays can be detected by the prototype telescope. The prototype will be setup in autumn 2012 in Berlin.

## INTRODUCTION

A prototype of a medium size telescope of the CTA (Cherenkov Telescope Array) project [1] is under preparation. It will be assembled in autumn 2012 in the science park Berlin-Adlershof (Germany). The prototype will not be fully functional, it has no photomultiplier in the camera housing and only a fraction of optical mirrors. The main goal is to prove the design concept and the conformity of the steel structure with the design. Furthermore a test of the drive system and a proof of the calibration methods are foreseen.  Measurements of static deformations and dynamic deformations depending on parameters of motion, angular position and environmental conditions will be performed.

## BASIC OPTICAL PARAMETERS

The focal length of the telescope 16 m, the diameter of the dish is 12 m and consequently the F/D is 1.35. With an active camera width of about 2.4 m the field of view is eight degrees. The 80% containment diameter of the optical Point Spread Function (PSF) out to 80% of the camera radius shall be smaller than the pixel size of 0.18 deg. The required precision of the orientation of the telescope axis during astrophysical tracking shall be better than 1% of the field of view diameter on each axis. This value represents roughly 5 arc minutes for the MST.

# PROTOTYPE MST-BASIC DESIGN

The tower is about 9 m in height and is supported by a concrete foundation. The tower supports the azimuth bearing, the azimuth drive system and the head as well. The yoke consisting of a right and left part is mounted on the head and holds the elevation bearings and drive systems in place. The counter weights are mounted on the rear side of the yoke to balance the combined weight of dish, quadrupod and camera. The dish is mounted onto the front side of the yoke. The dish will carry the 84 mirror facets.

The quadrupod both holds and supports the camera of about 2.5 tons in a distance of about 16 m from the center of the dish and is mounted onto the dish. The camera will not be instrumented, but has the same weight. The forces of wind and further environmental factors were considered in the design of the steel structure for different wind speeds.

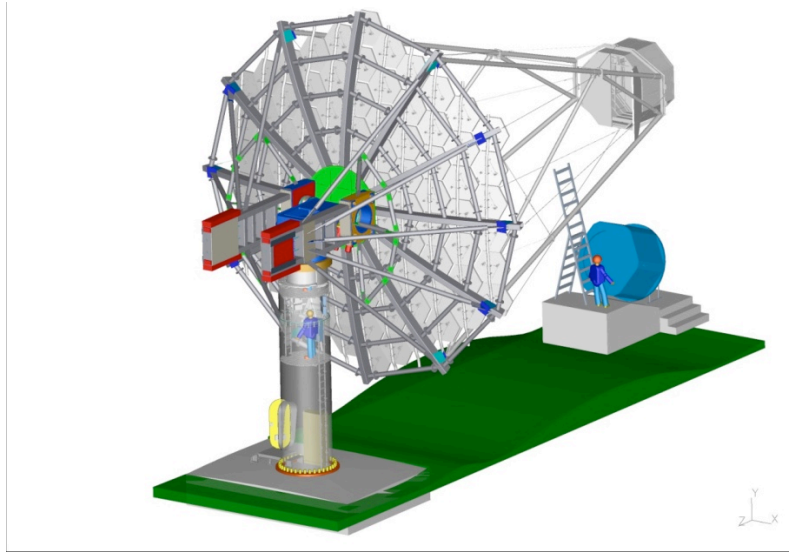

**FIGURE 1.** Design drawing of the Medium Size Telescope (MST) prototype

# DRIVE SYSTEM

The MST prototype design of the drive system has two independent axes: the azimuth axis and the elevation axis, one azimuth bearing and two elevation bearings. In order to have a high reliability and more flexibility, the azimuth axis is driven by two servo motors and the elevation axis by four servo motors.

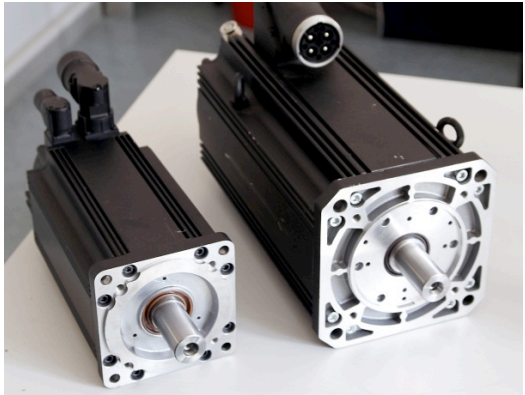

**FIGURE 2**  Motors for azimuth and elevation drive systems

## CALIBRATION AND POINTING

Calibration for the MST prototype means geometrical calibration, because the camera is not functional. For the first milestone a CCD-camera with a lens of 188 mm focal length is used. It is mounted in the dish off-axis by 5 m. Its optical axis will be aligned parallel to the dish axis with an accuracy of about 1 mrad using two optical refractive telescopes in combination with the CCD camera. Step by step the MST telescope will be moved so that bright stars and planets are in the sensor field of the pointing camera. The astronomical coordinates of these astronomical objects are well known. For each of these stars the internal incremental coordinates of the drive system are read and written into a table along with the astronomical coordinates. The estimated theoretical geometrical accuracy of the procedure is about pixel size/f = 4.5 microns / 188 mm = $2 \times 10^{-5}$. Further factors have to be considered which might increase the error: Accuracy of adjustment of the calibration setup, microscopic moving of the pointing camera, the uncertainty might presumably be larger than one pixel, uncertainty arising from the drive system like positioning accuracy, reproducibility etc. and uncertainties from possible deformations of the steel structure.

## SURVEY

The goal of the surveys is to prove the conformity of the steel construction compared to the design. Furthermore, the survey will allow to measure possible deformations depending on different angular positions and environmental parameters (wind, temperature). The methods are photogrammetry, computer-tachymetry and levelling.

## INSTRUMENTATION

The instrumentation is mainly based on CCD cameras. Five CCD cameras of unique type (Prosilica GC1350) will be used. Two are mounted in the center of the dish, one in the dish of axis and two in the dummy camera. In the near future also non-optical sensors are foreseen to be applied. The purpose for the CCD cameras are the measurement of the movement of camera dummy, furthermore two methods are foreseen for the mirror alignment, the pointing and calibration, the measurement and optimization of the PSF and the monitoring of parts of the dish and/or the position of the pointing camera.

## MEASUREMENT PROGRAM

The measurement program will start with the survey on the concrete foundation followed by the mounting of steel sub-assemblies. There is a stepwise sequence of survey parallel to steel structure mounting foreseen.
When dish and quadrupod are mounted, the start of the calibration procedure follows. After the mounting of the camera dummy the monitoring using CCD cameras to measure the movement of the camera dummy will start. The next task is the mounting of mirrors followed by mirror alignment using two different methods which have to be compared. A permanent survey will check that the steel structure is conforming to design including deformation checks depending on azimuth and elevation. A continuous monitoring by CCD cameras depending on parameters such as azimuth, elevation, PSF measurement and optimization, weather dependent deformation analysis will be performed. For the position monitoring of the dummy camera and the pointing camera LEDs are the light source. For all other measurements starlight is the light source.

## MIRRORS

The dish will carry 84 mirror facets. They are hexagonal and the width of the parallel edges is 1.2 m. They should have a low cost, low weight (< 20 kg), sufficient optical quality (PSF) and a long lifetime. The mirrors for the MST are produced by different institutes (CEA Saclay, INAF Brera and other institutes) based on similar technologies. Different coatings will be used, including multi-layers and di-electric coating. The mirror prototypes undergo an extensive test program including investigation of environmental factors.

### Mirror Mounting

A procedure for mounting the mirror facets is under development. It should not be time consuming and avoid any damage to the mirrors. One has to distinguish between mounting/dismounting of the full set of 84 mirror facets and mounting/dismounting of single mirrors or a small fraction of the full set whereas the main fraction of facets remains mounted. Corresponding to both cases the technologies and tools applied may differ. The system of catwalks on the rear side of the dish will be used for the mirror mounting and the commissioning and maintenance of the mirror control system (see below). Furthermore cranes and/or cherry pickers are used as well as different guiding facilities.

# Mirror Alignment

The goal of the prepared alignment procedures is the optimization of the mirror positions such, that the overall PSF of the telescope will be minimized. Two different procedures are in preparation: one similar to the HESS-like method elaborated at DESY in Zeuthen together with the University of Tuebingen. The other similar to the VERITAS method, elaborated at University of Zürich. CCD cameras are in both cases the basic tool. A simple image detection was developed for the HESS-like method [4]. From the image processing program the goal data in steps of the actuators moving the mirrors can be transferred to the control program which steers the mirrors. The linearity and reproducibility is checked. The accuracy is about 6 mm in the focal plane using lasers for the development of the method and a focal length of 30 mm for the CCD camera. Additional difficulties might arise from the fact that star light has to be used and the PSF will be folded to the optical signal. One has to consider that 84 spots from an equal number of mirrors will be projected onto the screen. An improvement could come from the application of optics of a larger focal length in a second step. This setup is foreseen to be used for the estimation and optimization of the PSF.

# SAFETY

The developed technical safety system has its main foci on personal safety and safety of the mechanical system. It considers all passive safety aspects, like lightning protection, safety fences and the mechanical latching of the camera, but also active safety aspects, like the use of limit switches and the implementation of an interlock system. A concept for personal safety is under development.

# ACKNOWLEDGMENTS

We gratefully acknowledge support from the agencies and organizations listed in this page: http://www.cta-observatory.org/?q=node/22.# REFERENCES

1. CTA consortium, "Design concepts for the Cherenkov Telescope Array CTA…", arXiv:1008.3703v3[astro-ph. IM], (11 April 2012).
2. E. Birsin et al., "Towards a Flexible Array Control and Operation Framework for CTA", Poster in *Proc. of the conference Gamma* , Heidelberg, Germany, (2012).
3. I.Oya et al., 2012, "Evaluating the Control Software for CTA in a MediumSize Telescope Prototype" in *Proc. of CHEP conference*, (2012).
4. Raeck, T., "Study of mirror prototypes for the Cherenkov Telescope Array project", Bachelor Thesis, Techn. Hochschule Wildau (FH), Germany, 2012.